# Automating Modelica Module Generation Using Large Language Models: A Case Study on Building Control Description Language

Hanlong Wan[a*], Xing Lu[a], Yan Chen[a*], Karthik Devaprasad[a], Laura Hinkle[a]

[a] Pacific Northwest National Laboratory, 3335 Innovation Blvd, Richland, WA 99354, USA
*Corresponding Author

Yan Chen (yan.chen@pnnl.gov), Hanlong Wan (hanlong.wan@pnnl.gov)

This research was supported by Battelle Memorial Institute under contract no. DE-AC05-76RL01830 with the U.S. Department of Energy (DOE). The publisher, by accepting the article for publication, acknowledges that the U.S. government retains a nonexclusive, paid-up, irrevocable, worldwide license to publish or reproduce the published form of this manuscript, or allow others to do so, for U.S. government purposes. DOE will provide public access to these results of federally sponsored research in accordance with the DOE Public Access Plan (http://energy.gov/downloads/doe-public-access-plan).

# HIGHLIGHTS

- Large language models were tested on four logic and five building control module tasks
- Claude-Sonnet-4 achieved a 100% success rate on logic blocks with the right prompts
- Control module generation achieved an 83% success rate with Claude
- Retrieval-augmented generation often failed; hard-rule search ensured correctness
- Productivity gains: 40%–60% time savings per module

# ABSTRACT

Dynamic energy systems and controls require sophisticated modeling frameworks to design and test advanced supervisory and fault-tolerant strategies. Modelica is a widely adopted, equation-based language, but developing control modules remains labor-intensive and requires specialized expertise. This paper investigates the potential of large language models (LLMs) to automate the generation of Control Description Language modules in Building Modelica Library as a case study. We designed a structured workflow that combines standardized prompt scaffolds, library-aware grounding, automated compilation with OpenModelica, and human-in-the-loop evaluation. Experiments were conducted across four basic logic tasks ("And," "Or," "Not," and "Switch") and five control modules (chiller enable/disable, bypass valve control, cooling-tower fan speed, plant requests, and relief-damper control). The results showed that GPT-4o failed to generate executable Modelica code in zero-shot mode, whereas Claude-Sonnet-4 achieved up to 100% success for basic logic blocks with carefully engineered prompts. For control modules, success rates reached 83%, and failed outputs required medium-level human repair (estimated 1–8 hours). Retrieval-augmented generation introduced frequent mismatches in selecting modules (e.g., "And" retrieved as "Or"), while a deterministic hard-rule search strategy eliminated these errors. Human evaluation also outperformed AI evaluation, as current LLMs cannot assess simulation results or validate behavioral correctness. Despite these limitations, the LLM-assisted workflow reduced the average development time from 10–20 hours down to 4–6 hours per module, translating to 40%–60% time savings. These findings underscore both the potential and current constraints of LLM-assisted Modelica generation, pointing toward future research in pre-simulation validation, improved grounding, and closed-loop evaluation.

# NOMENCLATURE

## ABBREVIATIONS

| | |
|---|---|
| AHU | air handling unit |
| API | application programming interface |
| ASHRAE | American Society of Heating, Refrigerating and Air-Conditioning Engineers |
| CDL | control description language |
| HVAC | heating, ventilating, and air-conditioning |
| LLM | large language model |
| MBL | Modelica Buildings library |
| RAG | retrieval-augmented generation |

## OTHER TERMS

| | |
|---|---|
| cooMod | cooling mode indicator |
| dpBui | building static pressure difference |
| Dymola | Dynamic Modeling Laboratory, a Modelica-based simulation tool |
| OpenModelica | open-source Modelica-based modeling and simulation environment |
| TChi_CHWST | chiller chilled-water supply temperature |
| TChiSet | chiller chilled-water temperature setpoint |
| TDeaBan | deadband for chiller enable/disable logic |
| VChiWat_flow | measured chilled water volume flow rate |
| VChiWatSet_flow | minimum chilled water volume flow rate setpoint |
| yValPos | bypass valve position signal |
| TCWSupSet | condenser water supply temperature setpoint |
| TCWSup | measured condenser water supply temperature |
| TCHWSupSet | chilled water supply temperature setpoint |
| TCHWSup | measured chilled water supply temperature |
| y | generic output signal |
| TAirSup | supply air temperature |
| TAirSupSet | supply air temperature setpoint |
| uChiWatPum | status of chilled water pump(s) |
| uCooCoi | cooling coil valve position signal |
| u1SupFan | supply fan status (Boolean) |
| yChiWatResReq | chilled water reset request signal |
| yChiPlaReq | chiller plant enable request signal |
| yRelDam | relief damper position signal |

# 1 INTRODUCTION

Dynamic energy systems including buildings, district heating and cooling networks, power systems, and renewable energy integration are central to improving efficiency, flexibility, and decarbonization across the energy sector. Effective supervisory and fault-tolerant controls can significantly reduce energy use, lower operating costs, and enhance grid reliability. For example, advanced controls have been shown to cut building energy consumption by up to 30% [1], with similar benefits demonstrated in district heating and multi-energy networks through optimized operation and predictive control. Despite this potential, deployment of advanced control strategies remains limited, leaving a substantial gap between technical capability and realized energy savings.

One barrier is the high cost and expertise required to design, implement, and test control sequences in practice. The Modelica Buildings library (MBL) is a library of simulation models for commonly used heating, ventilating, and air-conditioning (HVAC) components built with the Modelica language, an object-based acausal modeling language that allows users to create and simulate complex system models and their associated controls by assembling component models from predefined libraries [2]. The components in the MBL specifically facilitate the thermomechanical simulation of HVAC components. The Controls Description Language (CDL) is a language for specifying control sequences and algorithms by using block diagrams that can be first tested out in various simulation environments and then translated to machine code that is supported by various commercial control vendor platforms [3]. To facilitate simulation, the elementary blocks within the specification are implemented as a package of classes in the MBL. This allows users to both implement and test various control strategies within the Modelica environment by combining the elementary blocks to create control logic. As a part of the OpenBuildingControl project, standardized high-performance sequences for HVAC systems, as defined by American Society of Heating, Refrigerating and Air-Conditioning Engineers (ASHRAE) Guideline 36 ("High-Performance Sequences of Operation for HVAC Systems") [4], have been implemented through CDL and also released with the MBL [3].

Large language models (LLMs) are models that are trained on large text datasets and that use a probabilistic approach to generate new text on the basis of user prompts. The basis of LLMs is the transformer model, which is a neural network architecture that incorporates self-attention mechanisms to focus on the most relevant parts of the input [5]. For the prediction of new sequences of text, the textual input is first converted into a numerical representation via embeddings and then processed by the encoder-decoder architecture or a variant, which leverages self-attention mechanisms. There are several pretrained LLMs available for public use, including GPT [6], LLaMA [7], Claude [8], and BERT [9], among others. LLMs have been shown to be effective in writing code [10], and given the expertise required for writing CDL code, LLMs could be used to expedite the coding process.

Researchers have recently begun to explore the use of LLMs for generating Modelica code, aiming to reduce the amounts of expertise and time required for physics-based system modeling. Rupprecht et al. [11] showed that fine-tuning LLaMA on synthetic reactor scenarios improved the syntactic and semantic accuracy of Modelica models, though challenges remained in generalizing to unseen cases. Xiang et al. [12] proposed ModiGen—a workflow combining fine-tuning, retrieval-augmented generation (RAG), and feedback optimization—and demonstrated its significant improvements in simulation success and functional correctness across benchmark datasets. Complementing these structured approaches, Maxwell et al. [13] illustrated that even general-purpose LLMs like ChatGPT can generate approximate Modelica topologies for circuits or drivetrains, which can then be refined through repair strategies and parameter optimization. Collectively, these studies highlight both the potential and current limitations of LLMs in automating Modelica-based modeling, motivating further investigation into domain grounding, efficiency, and closed-loop validation.

This paper introduces a novel pipeline that integrates LLM-based CDL generation, automated compilation, simulation, and validation against reference libraries, especially for the building control domain. Unlike previously developed approaches, which primarily focus on natural language to code translation without feedback loops, our approach incorporates iterative evaluation, human-in-the-loop verification, and library-aware module selection. The main contributions are (1) demonstration of LLM capability in generating executable CDL modules for both basic logic and building control tasks, (2) quantitative evaluation of performance across different LLMs and prompt strategies, and (3) cost–benefit analysis of productivity gains in a real-world building control context. The remainder of this paper is organized as follows: Section 2 describes the methodology, Section 3 presents results, Section 4 discusses implications and limitations, and Section 5 concludes with an exploration of future research directions.

This study is the first to demonstrate LLM-assisted CDL module generation for building controls through the use of a structured pipeline with library-aware prompts and iterative validation. The workflow reduces expert effort by 40%–60% while ensuring standards-compliant, executable Modelica code, advancing practical AI integration in physics-based control modeling.

## 2     METHODS

### 2.1    Hierarchical modeling and LLM-based module generation in Modelica

Building control systems can be represented in Modelica using a hierarchical approach that progresses from basic logic to system-level control, as shown in Figure 1. At the lowest level, elementary operators such as "And," "Or," and "Not" provide the foundation for rule-based decision-making. These primitives are encapsulated into reusable control blocks that implement control sequences consistent with standards like ASHRAE Guideline 36, for use in equipment like dampers or economizers. For instance, damper modules determine actuator position according to fan status and outdoor air conditions, while economizer blocks regulate outdoor air intake to optimize energy efficiency.

At the system level, these component modules are assembled into complete HVAC control architectures. The MBL developed by Lawrence Berkeley National Laboratory provides preconfigured templates for single-zone and multizone air handling units, incorporating fans, coils, dampers, and supervisory controllers. This modular structure supports scalable modeling, validation against standard control sequences, and integration with building energy simulations, thereby enabling accurate representation of building-wide HVAC control strategies within a unified modeling environment.

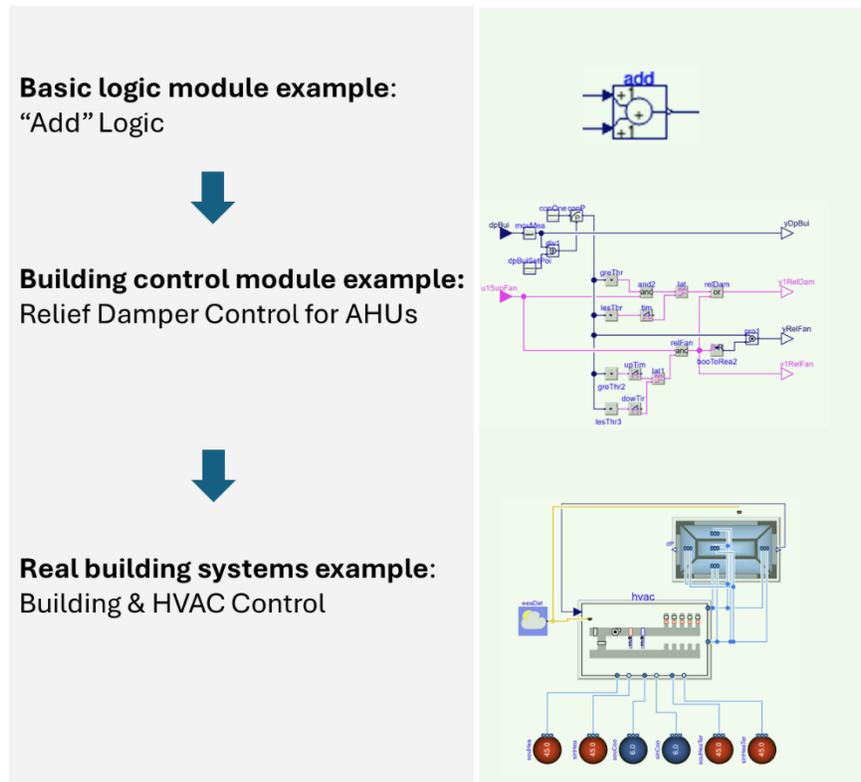

**Figure 1. Building control modules. AHU = air handling unit; HVAC = heating, ventilating, and air-conditioning.**

In this study, we focused on the generation of Modelica modules by using LLMs at the first two levels of abstraction: basic logic operators and component-level control modules. Once these foundational elements are reliably produced, the next step is to assemble them into more complex system-level configurations, such as whole-building HVAC control. Our methodology therefore establishes a structured starting point for automated model generation, where LLMs assist in producing validated submodules that can subsequently be composed into integrated, large-scale building control systems.

### 2.2 Workflow

The first stage of our workflow addresses the generation of basic logic modules, which serve as foundational building blocks for higher-level control. These modules (e.g., "And," "Or," "Not," and "Switch") are generated directly from user requests through zero-shot prompting or lightly structured prompts. We systematically evaluated the influence of different LLMs (including Claude-Sonnet-4 and GPT-4o) as well as the effect of prompt design on generation outcomes. We assessed performance across four case study modules by comparing compilation success rates, token consumption, and response time. This allowed us to quantify the trade-offs between LLM families and prompting strategies in terms of reliability and computational efficiency.

The second stage focuses on building control module generation, where the workflow follows a structured process (Figure 2). After user tasks are decomposed into submodules, related components are identified either through a hard-rule search within the CDL root or via an RAG approach. Hard-rule searches constrain references strictly to existing library modules, offering improved reliability compared with that of unconstrained retrieval. Draft Modelica modules are then generated by the LLM and iteratively validated through the OpenModelica application programming interface (API). Errors in loading or simulation trigger correction loops, where additional LLM prompts refine the code until convergence. Beyond syntactic correctness, we also explored the ability of LLMs to assess behavioral performance by comparing generated modules against human-prepared reference models and intended task requirements, offering a first step toward automated self-evaluation of Modelica scripts.

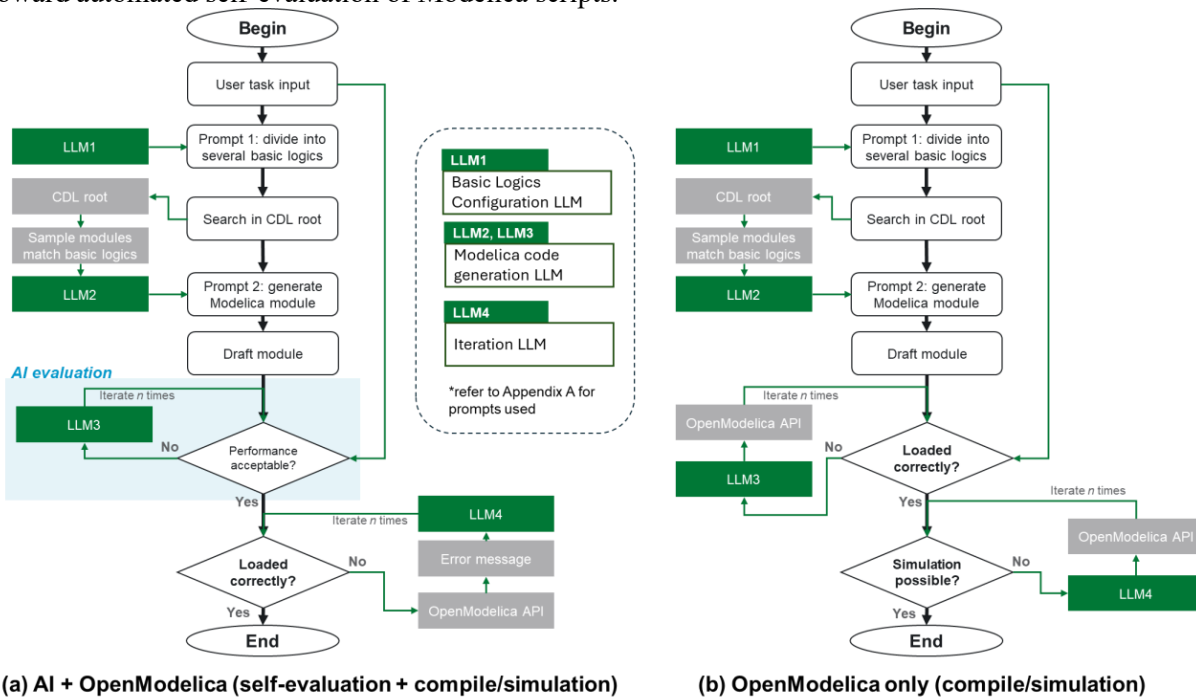

**Figure 2. Workflow of building control module generation (a) with AI evaluation in-the-loop (b) without AI evaluation. LLM = large language model; CDL = control description language; API = application programming interface.**

In our framework, we designed three specialized LLM roles, each serving a distinct purpose in the pipeline. Model 1, the Modelica code expert, focuses exclusively on generating Modelica modules (LLM2, LLM3, LLM4 in Figure 2). Model 2, the building control expert, interprets user queries and translates them into appropriate control logics, selecting from existing module structures (LLM1 in Figure 2). Model 3, the evaluation expert, is responsible for assessing both the correctness of generated Modelica modules and the performance of simulation outputs relative to the task requirements (not shown in Figure 2, refer to section 4). To ensure robustness and correctness, we implemented three iterative loops. The compilation loop employs the OpenModelica API to load generated modules, with LLM-guided refinement used to resolve syntax or structural errors. The simulation loop also leverages the OpenModelica API to execute the models, iterating with LLM assistance to address runtime issues. Finally, the evaluation loop relies entirely on the LLM to analyze generated modules and simulation outputs against the specified task objectives, enabling a self-correcting and task-aware evaluation process. Together, these role-based experts and iteration loops create a structured workflow for automated generation, execution, and assessment of control-oriented Modelica modules.

### 2.3 Test example design

We standardized prompt design to produce library-conformant CDL/Modelica controllers. Every prompt follows the same three-step template:

1. A goal sentence that explicitly asks the model to "generate a Modelica control block for …"
2. An interface declaration with named inputs and outputs (including types, where relevant)

3. A concise description of the control sequence as rules or state logic aligned with the MBL

We kept the phrasing uniform ("Please generate … The inputs are … The outputs are … The control sequence is …"), so tasks differ only in content, not in form.

This template was applied across waterside and air-side controllers, and it maps directly to canonical MBL applications. Representative cases (as listed in Table 1) include chiller enable/disable using chilled-water leaving temperature with a deadband; a chilled-water minimum-flow bypass valve per ASHRAE Guideline 36 §5.20.8; cooling-tower fan speed with mode-dependent objectives; plant-level request signals for chilled-water reset and chiller plant enable per G36 §5.16.16; and relief-damper control that maintains building static pressure with a P-only loop when the supply fan is proven to be on.

**Table 1. Synopsis of prompt examples and coverage.**

| Task | Control block | Goal phrase used | Inputs | Outputs | Core sequence logic |
|---|---|---|---|---|---|
| 1 | Chiller enable/disable | "generate a Modelica control block for enabling and disabling a chiller" | TChi_CHWST, TChiSet | y | Enable if TChi_CHWST > TChiSet + TDeaBan; disable if TChi_CHWST ≤ TChiSet |
| 2 | CHW minimum-flow bypass valve (G36 §5.20.8) | "generate a block for the … minimum-flow bypass valve controller" | VChiWat_flow, VChiWatSet_flow, uChiWatPum | yValPos | If any CHW pump is proven to be on, enable the Proportional–integral–derivative (PID) loop; otherwise, drive the valve 100% open; set the bias to 100% when enabled |
| 3 | Cooling-tower fan speed | "generate a … cooling-tower speed control" | TCWSupSet, TCWSup, TCHWSupSet, TCHWSup, cooMod | y | Full mechanical cooling: track CWST to setpoint; Part mechanical cooling: 100% speed; Free cooling: track CHWST to setpoint |
| 4 | Plant requests (G36 §5.16.16) | "generate … for chilled-water reset request and chiller plant request" | TAirSup, TAirSupSet, uCooCoi | yChiWatResReq, yChiPlaReq | Tiered requests based on SAT offset and coil valve position with hysteresis thresholds |
| 5 | Relief-damper control (G36 §5.16.8) | "generate … relief-damper control … without fan" | dpBui, u1SupFan (Bool) | yRelDam | Enable when the supply fan is proven to be on; activate the P-only loop to maintain the dp setpoint; close when disabled |

## 2.4 Human validation

We assessed every generated controller with a standardized human-evaluation workflow designed to separate fundamental behavioral correctness from secondary qualities of implementation, as shown in Table 2.

**Table 2. Template for human evaluation.**

| Modelica Control Block Evaluation Template |||||
|---|---|---|---|---|
| **Section 1: Task Compliance Check** |||||
| Question || Response (Yes/No) || Notes |
| Does the module behave correctly for the control task? <br> ➙ If "Yes," proceed to Section 2 <br> ➙ If "No," skip Section 2 and complete Section 3 instead |||||
| **Section 2: Scoring Path A (If Task Behavior Is Correct)** |||||
| **Criterion** | **Description** || **Score: 0 or 1 (1 for yes; 0 for no)** | **Comments** |
| Library Appropriateness | Only uses the expected library (e.g., Buildings.Controls.OBC.CDL.*), not general Modelica blocks |||||
| Structure / Modularity / Readability | The model is neatly organized, modular, and understandable |||
| Interface Accuracy | Inputs/outputs are clearly named, correctly typed, and documented with appropriate units |||
| Logic Simplicity / Clarity | Control logic is written in a simple, understandable, and efficient way |||
| Robustness (e.g., Hysteresis) | Includes measures to prevent short cycling (e.g., latch or hysteresis) |||
| **Score** |||||
| **Section 3: Scoring Path B (If Task Behavior Is NOT Correct)** |||||
| **Criterion** | **Description** || **Score: 0 or 1 (1 for yes; 0 for no)** | **Comments** |
| Simulation & Syntax Validity | It can parse and compile without grammar/syntax errors |||
| Semantic Correctness | Uses correct types, legal Modelica declarations, and proper connections |||
| Logical Soundness | Logic makes internal sense even if it doesn't meet the control task |||
| Interface Appropriateness | Inputs/outputs correctly are typed, named, and reasonably documented |||
| **Score** |||||
| **Section 4: Time Required to Correct the Model** |||||
| **Number** | **Problem** || **Efforts to correct the module** | **Comments** |
|  |  |  |  |  |

The reviewer first performs the task-compliance gate by evaluating the module's observable behavior relative to the intended control logic described in the prompt. For example, a relief-damper controller is considered compliant if it opens only when the supply fan is proven on, modulates to maintain the building's static pressure at its setpoint, and remains closed when the fan is off; any inversion of conditions or omission of the fan-proof interlock constitutes non-compliance. If the gate is passed, the reviewer proceeds to quality scoring on the basis of five binary criteria:

- **Library appropriateness:** Reflects whether the implementation draws from the expected MBL scope rather than ad hoc compositions that undermine reuse.
- **Structure, modularity, and readability:** Indicates whether the design is decomposed into coherent sub-blocks (e.g., setpoint generation, enable/disable, and actuator loop) with clear dataflow.
- **Interface accuracy:** Verifies that inputs and outputs are correctly named, typed, and, where applicable, carry units and sensible default parameterization.
- **Logic simplicity and clarity:** Rewards direct expression of the control policy, for instance, tiered requests with explicit thresholds and hysteresis, while penalizing redundant or unreachable branches.
- **Robustness:** Indicates whether the design incorporates deadbands, hysteresis, latching, or minimum on/off times to prevent short cycling; for example, a chiller-enable model that compares chilled-water leaving temperature to a setpoint with a deadband is considered robust, whereas a single-threshold comparator that chatters is not.

When the gate is not passed, the reviewer provides a concise diagnosis rather than scoring the quality. Typical fault classes include duplicated control elements (such as an extra PID path feeding the same actuator), inverted inequalities that reverse actuator intent (e.g., a minimum-flow bypass valve driven closed when measured flow is below the minimum setpoint), nonexistent or mismatched submodule references, and broken connections that keep signals from propagating. The reviewer also records a remediation estimate by using qualitative effort bands. Minor effort denotes quick edits (e.g., deleting a duplicate block or correcting a connector name). Moderate effort covers refactoring a logic branch or repairing thresholds and latches. Major effort indicates substantial redesign or rewiring, such as when essential components are missing or incompatible. These records, together with brief justifications, create a fault catalog that is useful for method comparison and for guiding subsequent model repair within the compile/simulation/evaluation loops.

All decisions and short rationales are recorded for each controller and reviewer. For controllers that pass the gate, the quality score is the mean score across the five binary criteria; we then report the average and dispersion of these means across reviewers and tasks, together with per-criterion pass rates, to highlight recurring strengths and weaknesses. For noncompliant controllers, we summarize the distribution of fault types and remediation bands. When multiple reviewers assess the same artifact, discrepancies at the gate or on any criterion are reconciled through brief adjudication using the written comments and the controller diagram, after which a consensus record is stored.

## 3   RESULTS

### 3.1   Basic logic generation

In basic logic module generation, we first attempted to run a smaller open-source model locally (LLaMA-3 8B), but all trials failed to compile. We then moved on to larger commercial models. As shown in Figure 3., Claude-Sonnet-4 achieved a 100% compilation success rate, while GPT-4o consistently failed in zero-shot mode. The response time distributions indicate that GPT-4o was the fastest (typically <1 s), whereas LLaMA-3 8B averaged ~1.5–2 s and Claude required ~2–3 s. Token count distributions further highlight model differences: Claude consistently produced higher token counts (~170–180) compared with GPT-4o (~90–140) and LLaMA-3 8B (~20–70). We considered token count because it directly correlates with cost, making it a key metric for comparing models. Claude's higher token count does not reflect wasted tokens— it reflects Claude's ability to generate meaningful, structured outputs and successfully compile them into executable Modelica code. Overall, these findings confirm Claude's superior performance for domain-specific tasks, while also emphasizing the trade-offs between runtime efficiency, generation cost, and syntactic correctness. Our finding on the model performance of Modelica is consistent with the findings reported by Xiang et al. [12].

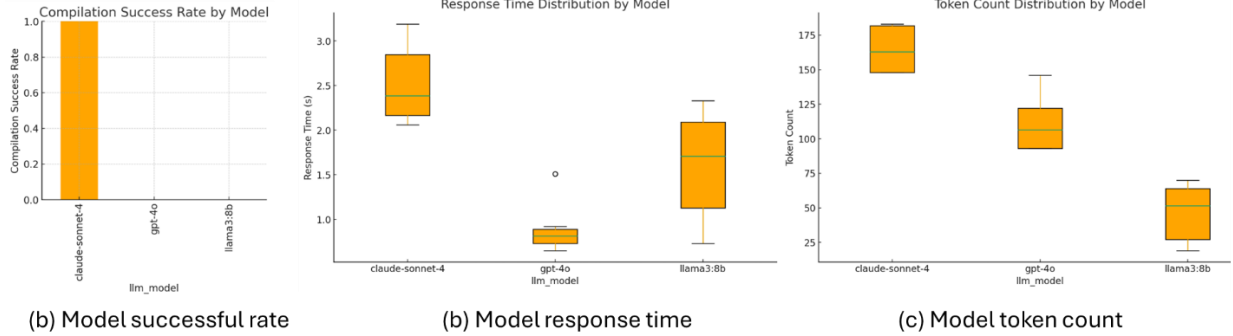

Figure 3. Comparison of generation performance between Claude and GPT.

Next, we attempted to use Claude to generate several typical basic logic modules, as shown in Figure 4. The generation prompts are provided in Appendix A – Prompts used. We discovered that the prompt has a strong impact on the module generation. For the "Switch" module in particular, which is less commonly used than the others, the success rate can be very different. With the right prompts, the success rate for basic logic generation can reach almost 100%.

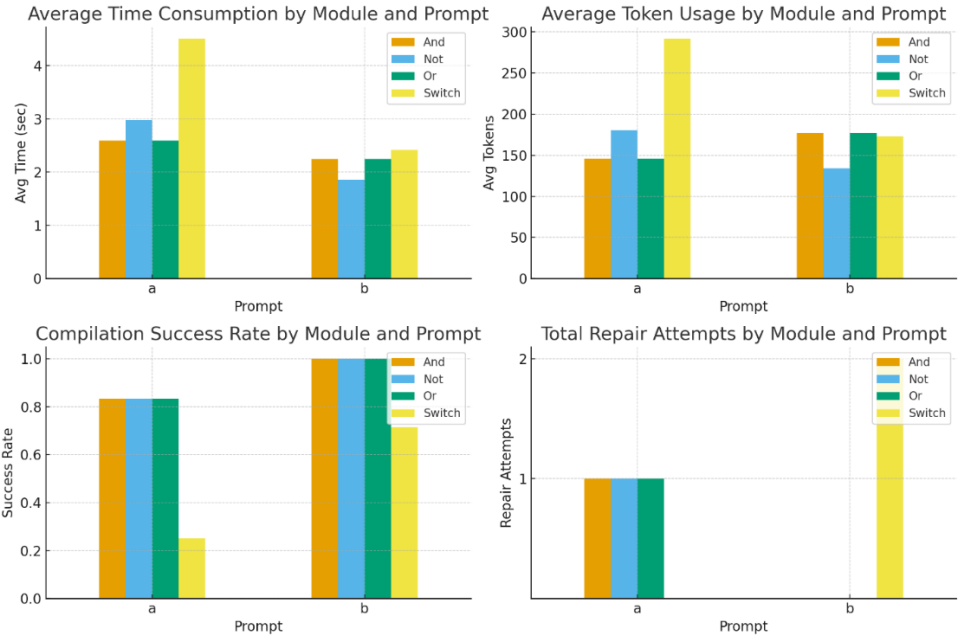

**Figure 4. Generation performance for four typical modules.**

### 3.2 Basic control module generation

In our initial experiments, we attempted "zero-shot" prompting for CDL module generation, but this approach failed consistently across models. As summarized in Table 3, only Claude produced partially correct outputs, achieving syntactic validity in some cases but still struggling with semantic accuracy and structural integrity. GPT-4o and GPT-4o-mini, by contrast, failed entirely in zero-shot mode. In response to these results, we discontinued zero-shot prompting and instead moved forward with library-aware prompting strategies, where the CDL library was explicitly provided as context. All subsequent experiments therefore focused on Claude, which showed the most promise among the tested models.

**Table 3. "Zero-shot" prompt generation for test examples.**

| Task | Claude-4 | | | | GPT-o4 | | | | GPT-o4mini | | | |
|---|---|---|---|---|---|---|---|---|---|---|---|---|
| | Syntax | Semantic | Structure | Work? | Syntax | Semantic | Structure | Work? | Syntax | Semantic | Structure | Work? |
| 1 | ✓ | ✓ | X | ✓ | ✓ | ✓ | X | X | ✓ | X | X | X |
| 2 | ✓ | X | ✓ | X | X | X | X | X | ✓ | ✓ | X | X |
| 3 | ✓ | X | ✓ | X | ✓ | ✓ | X | X | X | X | X | X |
| 4 | ✓ | ✓ | X | X | ✓ | X | X | X | ✓ | X | X | X |
| 5 | ✓ | X | ✓ | X | X | X | X | X | ✓ | ✓ | ✓ | ✓ |

#### 3.2.1 RAG vs. Hard-rule search

We initially explored an RAG pipeline to guide the LLM in selecting appropriate CDL submodules. However, this approach often introduced errors due to semantic overlap in library text. For example, when tasked with retrieving the "And" module, the RAG method sometimes returned the "Or" module instead because its documentation contained references to "and." To eliminate such mismatches, we adopted a hard-rule search strategy. In this approach, the contents of the CDL library were indexed explicitly, and only exact matches to existing module names were pulled, ensuring reliable grounding in the library without false positives.

### 3.2.2 Human evaluation vs. AI evaluation

**Table 4. Human evaluation vs. AI evaluation.**

|  |  | Task 1 | Task 2 | Task 3 | Task 4 | Task 5 |
|---|---|---|---|---|---|---|
| Workflow 1 (**without AI** evaluation) | Human Expert 1 | Yes | No | No | Yes | No |
|  | Human Expert 2 | Yes | Yes | Yes | Yes | No |
|  | AI evaluation | No | N.A. | Yes | No | No |
| Workflow 2 (**with AI** evaluation) | Human Expert 1 | No | No | No | No | No |
|  | Human Expert 2 | No | Yes | Yes | N.A. | N.A. |
|  | AI evaluation | No | N.A. | No | No | Yes |

Using the workflow shown in Figure 2(a), we compared outcomes with and without AI-based evaluation. As summarized in Table 4, to test evaluation capability, we implemented two pathways: (1) sending simulation results as time-series data to the LLM and (2) directly sending the generated Modelica module together with the initial task request. The results showed that although AI can reliably detect compilation errors, its ability to assess behavioral correctness remains limited. Human experts consistently outperformed AI evaluation across tasks, underscoring that current LLMs cannot yet autonomously improve module design or verify performance beyond syntax checks. Nonetheless, incorporating AI evaluation into the workflow provides an additional screening layer that reduces trivial errors and highlights the potential for more advanced, closed-loop validation in future work.

### 3.2.3 Final performance

Table 5 shows the final workflow performance. Across all test cases, the overall success rate of the LLM-assisted pipeline reached 83%, with the majority of generated modules compiling and simulating without major intervention. For the remaining cases, although the modules were not immediately correct, the required effort to repair them was consistently in the medium range (1–8 hours). This level of rework is still substantially lower than the baseline effort level estimated by domain experts: Expert 1 stated that developing comparable modules entirely from scratch would take 6–8 hours, and Expert 2 estimated that the process would take 8–12 hours. Thus, even in instances where corrections were necessary, the AI-assisted workflow reduced total development burden.

**Table 5. Final success rate and level of effort required to correct the modules.**

|  | Task 1 | Task 2 | Task 3 | Task 4 | Task 5 |
|---|---|---|---|---|---|
| Human Expert 1 |  |  |  |  |  |
| Trial 1 | Yes | Yes | Yes | No *Medium* | Yes |
| Trial 2 | Yes | Yes | No *Medium* | Yes | Yes |
| Trial 3 | Yes | Yes | No *Medium* | Yes | Yes |
| Trial 4 | Yes | Yes | Yes | Yes | Yes |
| Human Expert 2 |  |  |  |  |  |
| Trial 1 | Yes | Yes | Yes | No *Medium* | Yes |
| Trial 2 | Yes | Yes | No *Medium* | Yes | Yes |

Overall, productivity analysis confirmed that LLMs can reduce development time by 40%–60%, from 10–20 hours down to 4–6 hours per model. At an assumed labor rate of $100/hour, this translates to $600–$1,600 in savings per model, or approximately $30,000–$160,000 across 50–100 models. Simulation-based validation further demonstrated that structurally correct modules generally functioned as intended, although approximately 20% required additional human debugging to fully align with task-specific control requirements. Taken together, these findings demonstrate both the practical reliability and economic value of integrating LLM-assisted workflows into CDL-based Modelica module development.

## 4 DISCUSSIONS

### 4.1 Model performance

The evaluation results demonstrate that among the tested LLMs, those in the Claude family have the strongest capability for generating executable CDL modules. In particular, Claude-Sonnet-4 consistently produced syntactically correct code and, with carefully designed prompts, achieved a success rate of up to 83%. By contrast, GPT-4o and GPT-4o-mini failed for almost all zero-shot prompting scenarios. These findings underscore the critical role of prompt engineering in achieving reliable outcomes, as less common modules (e.g., "Switch") were particularly sensitive to prompt design.

Nevertheless, limitations remain evident. As of mid-2025, Claude models lack sufficient domain-specific reasoning to evaluate building-oriented Modelica modules or interpret simulation results from OpenModelica/Dymola. Attempts to use Claude for simulation analysis or correction suggestions were largely unsuccessful, confirming that human oversight is still required in the evaluation loop. Furthermore, iteration between simulation and LLM correction is slowed by repeated library loading and runtime overheads, limiting the responsiveness of the workflow. Despite these challenges, the productivity gains are notable: LLM assistance reduced development time from 10–20 hours to 4–6 hours per model, representing 40%–60% time savings compared with expert-only development.

### 4.2 Pipeline comparisons

Two strategies were tested for grounding module generation in the CDL library: RAG and hard-rule submodule search. The RAG approach, while conceptually flexible, frequently introduced retrieval errors due to semantic overlaps in library documentation. For instance, when tasked to retrieve the "And" block, RAG occasionally returned the "Or" module because the description for this module contained the word "and." Such errors compromised the reliability of the pipeline.

By contrast, the hard-rule search strategy ensured that only exact matches to module names present in the CDL library were retrieved. While less flexible for generating novel or unseen control sequences, this deterministic approach guaranteed compatibility and correctness for library-based modeling tasks. A strong advantage of this approach is that LLM-generated modules can be restricted to a specific library, such as the OpenModelica Buildings library, without drifting into the broader Modelica Standard Library (MSL). This constraint is particularly valuable for developing standard-compliant control logic (e.g., ASHRAE Guideline 36), where adherence to library conventions is essential.

## 4.3 Lessons learned from human evaluation

We observed functional equivalence but differences in block granularity when comparing AI-generated designs with hand-built implementation from the MBL library. Figure 5 shows detailed sample comparisons. Content generated in Task 4 and human-prepared content can be found in Appendix B Sample Modules.

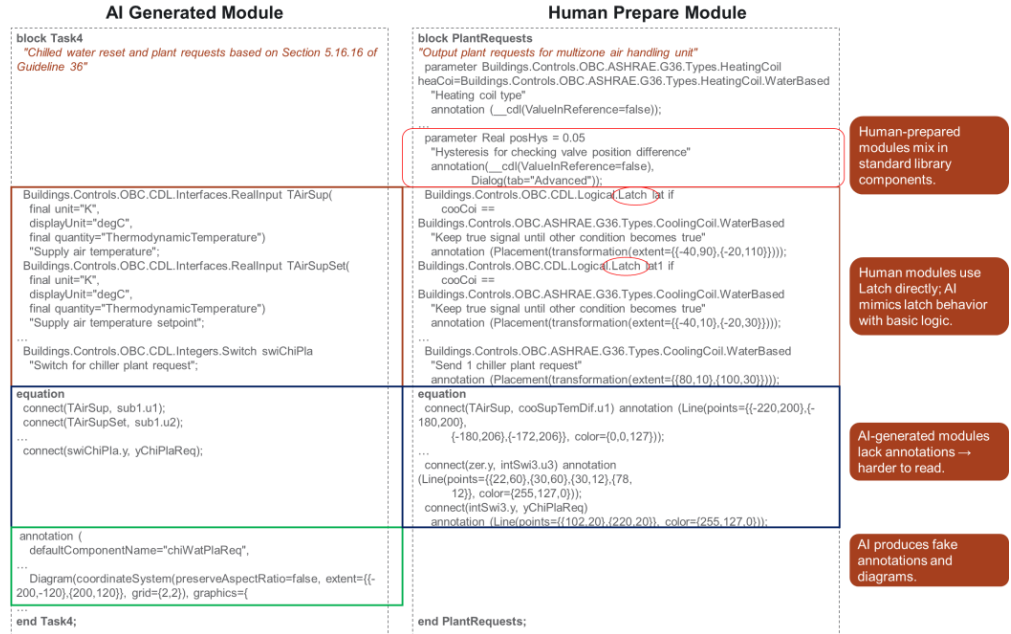

**Figure 5. Comparisons between the AI-generated module and human-prepared module.**

In Task 4, the manual example involved computation of the chilled-water reset request with "GreaterThreshold," "LessThreshold," and "Latch." The AI example involved the same action, but with the assembly of lower-level blocks such as "add," "or," and "hysteresis." A plausible cause is that the model favors widely applicable primitives that were prevalent in its training data and is less aware of library-specific composite patterns that improve reuse and readability. Prompt phrasing that emphasizes rule lists may further steer the model toward primitive logic assembly. This can be improved by retrieving and few-shotting canonical MBL patterns, adding a reward for a smaller block count or higher reuse, and applying rewrite rules that collapse clusters of primitives into standard threshold or latch constructs.

We saw frequent errors in controller action direction. In Task 2, the PI controller was designed to keep the valve open until the minimum flow had been met, yet the generated logic inverted the action. In Task 4, the proportional controller for relief dampers was designed to open when the building pressure had risen above the setpoint, but the sign was flipped. The cause was likely ambiguity in sign conventions, actuator polarity, and unit semantics that are not explicit at the interface. Correctness often requires domain knowledge about whether an increasing control signal opens or closes the device. A way to address this is to make polarity explicit in the interface, provide a small input–output trace that shows the desired monotonic response, add unit and direction checks, and run a short closed-loop simulation that verifies that $\partial$output/$\partial$error has the intended sign under typical conditions.

We found version-related inconsistencies due to changes across library releases. Buildings 9.0.x and 10.1.x differ in package structure (e.g., Buildings.Controls.OBC.CDL.Continuous.Sources in 9.0.x vs. Buildings.Controls.OBC.CDL.Reals.Sources in 10.1.x) and parameter availability. The AI sometimes mixed components across versions or referenced parameters that no longer exist. Similar issues arose with the Modelica Standard Library. The root cause was that the generator lacked reliable knowledge of the target environment at synthesis time. Mitigation included pinning the exact MBL and MSL versions in both the prompt and toolchain, performing a preflight import to expose available classes and parameters to the model, maintaining a compatibility map that rewrites deprecated names, and validating the generated Modelica/CDL modules by compiling them with the pinned toolchain during the generation loop.

Diagram layout and icon annotations remain out of scope in the current framework. Some models can produce annotations, but the results are inconsistent and difficult to read. A practical path forward is to separate code generation from presentation. First generate correct and validated CDL or Modelica code. Next, apply a post-processing stage to render diagrams by using a fixed icon set from the target library, an automatic layout based on the connection graph, and a minimal annotation policy that enforces legible labels and non-overlapping connectors. Future evaluations can score readability and consistency independently from functional correctness.

### 4.4 Future work

Looking forward, we can identify several areas for improvement. First, pre-simulation correctness checks could allow future LLMs to assess Modelica code validity prior to compilation, reducing the costs resulting from the compile–debug–repair cycle. Second, automated results interpretation remains a critical gap: More advanced models could analyze time-series simulation outputs and verify alignment with specified control requirements, thereby eliminating the current reliance on human evaluators. Third, closed-loop validation frameworks may eventually become feasible, where LLMs not only generate but also iteratively validate and refine modules with minimal human intervention. Finally, pipeline efficiency—particularly reducing library loading overhead and accelerating simulation-LLM iteration—remains an important practical consideration for real-world adoption.

Taken together, these findings highlight both the current strengths of Claude-based generation in producing executable CDL code and the limitations that necessitate human-in-the-loop oversight. While today's workflows deliver measurable productivity gains, future advances in LLM reasoning and validation could enable a more autonomous, efficient, and trustworthy pipeline for building control module development.

## 5    CONCLUSION

This study evaluated the potential of LLMs to automate the generation of Modelica-based control modules. The results demonstrate that for four representative logic tasks, Claude-Sonnet-4 achieved up to 100% success with carefully engineered prompts, whereas GPT-4o failed consistently in zero-shot mode. Extending the evaluation to five building control tasks, the overall pipeline attained a best-case success rate of 83%, although the remaining percent (17%) of the modules required human debugging to correct logic errors or submodule mismatches.

Two strategies for library grounding were compared: RAG and hard-rule search. While RAG frequently mis-selected modules (e.g., confusing "And" with "Or"), the deterministic hard-rule search eliminated such errors entirely. Furthermore, attempts to incorporate AI-based evaluation revealed that while LLMs could detect syntax errors, they lacked the capability to verify simulation correctness, leaving human evaluators indispensable for assessing functional behavior.

Despite these limitations, the LLM-assisted workflow produced tangible productivity gains. The amount of required expert development effort was reduced from 10–20 hours to 4–6 hours per module, corresponding to 40%–60% time savings. At an assumed labor rate of $100 per hour, this translates into a potential cost reduction of $600–$1,600 per module, or $30,000–$160,000 across 50–100 modules. These findings highlight both the promise and current boundaries of AI-assisted modeling, indicating that while human

oversight remains critical, structured LLM workflows can already accelerate the development of building control modules.

## AUTHORSHIP CONTRIBUTION STATEMENT

**Hanlong Wan**: Conceptualization, methodology, validation, formal analysis, software, writing (original draft), visualization, and investigation. **Xing Lu**: Validation, formal analysis, writing (original draft and human validation), visualization, and investigation. **Yan Chen**: Supervision, funding acquisition, writing (review and editing), and project administration. **Karthik Devaprasad**: Validation of testing modules and writing (original draft, literature review, and control module). **Laura Hinkle**: Writing (original draft, literature review, and large language model), and investigation (basic logic generation).

## DECLARATION OF COMPETING INTERESTS

The authors declare that they have no known competing financial interests or personal relationships that could have appeared to influence the work reported in this paper.

## ACKNOWLEDGMENTS

The research described herein was funded by the Generative AI for Science, Energy, and Security Science & Technology Investment under the Laboratory Directed Research and Development Program at Pacific Northwest National Laboratory (PNNL), a multiprogram national laboratory operated by Battelle for the U.S. Department of Energy. This work was also supported by the Center for AI and Center for Continuum Computing at PNNL. The authors would also like to acknowledge Victoria R. Scanlon for her assistance with copyediting.

# CODE AND DATA AVAILABILITY

The implementation of the proposed workflow is provided in a public repository at https://github.com/pnnl/prompt2control.

# APPENDIX A – PROMPTS USED

A. **Prompts used for basic logic generation**

1. AND
   a. "Write a Modelica component that implements a parameterized logical AND block. Use only the Modelica Standard Library, and return only Modelica code. Do not include comments or explanations."
   b. "Write a Modelica component that implements a parameterized logical AND block. The block should: - Be named `And`. - Accept an array of Boolean inputs as a parameterized input. - Output a single Boolean y, which is the logical AND of all inputs in the array. Use only the Modelica Standard Library, and return only Modelica code. Do not include comments or explanations."
2. OR
   a. "Write a Modelica component that implements a parameterized logical OR block. Use only the Modelica Standard Library, and return only Modelica code. Do not include comments or explanations."
   b. "Write a Modelica component that implements a parameterized logical OR block. The block should: - Be named `Or`. - Accept an array of Boolean inputs as a parameterized input. - Output a single Boolean y, which is the logical OR of all inputs in the array. Use only the Modelica Standard Library, and return only Modelica code. Do not include comments or explanations."
3. NOT
   a. "Write a Modelica component that implements a parameterized logical NOT block. Use only the Modelica Standard Library, and return only Modelica code. Do not include comments or explanations."
   b. "Write a Modelica component that implements a parameterized logical NOT block. The block should: - Be named `Not`. - Accept a single Boolean input as a parameterized input. - Output a single Boolean y, which is the logical negation of the input. Use only the Modelica Standard Library, and return only Modelica code. Do not include comments or explanations."
4. SWITCH
   a. "Write a Modelica component that implements a parameterized logical SWITCH block. Use only the Modelica Standard Library, and return only Modelica code. Do not include comments or explanations."
   b. "Write a Modelica component that implements a parameterized logical SWITCH block. The block should: - Be named `Switch`. - Accept three Boolean inputs as a parameterized input. The first input is the first input signal. The second input is the Boolean switch that determines which input signal is passed to the output. The third input is the second input signal. - Output a single Boolean y. Use only the Modelica Standard Library, and return only Modelica code. Do not include comments or explanations."

B. **Prompts used for control module generation**

1. Modelica code generation LLM
   System prompt
   "You are a code generator. Only return valid Modelica code with no natural language, no explanations, and no comments.

   Follow these additional modeling and layout conventions:

- Use only modules from Buildings.Controls.OBC.CDL or the Modelica Standard Library.
- For any temperature-related `input` or `parameter`, use:
`final unit="K", displayUnit="degC", final quantity="ThermodynamicTemperature"`.
- For Boolean or normalized output signals (e.g., y = 0 or 1), use:
`final min=0, final max=1, final unit="1"`.

Graphic layout guidance:
- Use the order of `connect()` statements in the `equation` section to infer logic flow: upstream instances appear earlier as sources, downstream ones appear later as targets.
- Reflect this left-to-right logic flow in the `annotation` section by placing upstream components to the left and downstream components to the right.
- Ensure all instances are positioned with unique `(x, y)` coordinates to avoid overlapping.
- Use clear spacing between instances to improve readability in the `diagram` layer.
- Annotate instances with their names where appropriate to enhance visual understanding."

Input Prompt
"Using these relevant CDL modules:{modules}
and the control task:{task}
Generate the Modelica code block that fulfills the control task."

2. Basic Logics Configuration LLM
System Prompt
"You are an expert Builidng Control engineer. Respond in structured, clear text."

Input Prompt
"Given the following control task, and available CDL module name, list the **CDL modules** you would use in bullet
points with no explanations, no comments. As few as possible CDL modules should be used to achieve the task.
Task:
{task}
available CDL modules:
{txt}"

3. Iteration LLM
System Prompt
"You are an expert to evaluate Modelica models. You can only answer yes or no."

Input Prompt
"The following Modelica code has compilation errors:
{error_log}
Please suggest a corrected version of the code.
Code: {code_content}
Corrected Modelica code:"

# APPENDIX B SAMPLE MODULES

## Task 4 – AI generated

```
within Buildings.Controls.OBC.CDL.Examples;
block Task4
  "Chilled water reset and plant requests based on Section 5.16.16 of Guideline 36"

  Buildings.Controls.OBC.CDL.Interfaces.RealInput TAirSup(
    final unit="K",
    displayUnit="degC",
    final quantity="ThermodynamicTemperature")
    "Supply air temperature";

  Buildings.Controls.OBC.CDL.Interfaces.RealInput TAirSupSet(
    final unit="K",
    displayUnit="degC",
    final quantity="ThermodynamicTemperature")
    "Supply air temperature setpoint";

  Buildings.Controls.OBC.CDL.Interfaces.RealInput uCooCoi(
    final min=0,
    final max=1,
    final unit="1")
    "Chilled water valve position";

  Buildings.Controls.OBC.CDL.Interfaces.IntegerOutput yChiWatResReq
    "Chilled water reset request";

  Buildings.Controls.OBC.CDL.Interfaces.IntegerOutput yChiPlaReq
    "Chiller plant request";

  Buildings.Controls.OBC.CDL.Reals.Subtract sub1
    "Temperature difference";

  Buildings.Controls.OBC.CDL.Reals.GreaterThreshold greThe3K(
    final t=3)
    "Check if temperature difference is greater than 3K";

  Buildings.Controls.OBC.CDL.Reals.GreaterThreshold greThe2K(
    final t=2)
    "Check if temperature difference is greater than 2K";

  Buildings.Controls.OBC.CDL.Logical.TrueDelay truDel3K(
    final delayTime=120)
    "2 minute delay for 3K difference";

  Buildings.Controls.OBC.CDL.Logical.TrueDelay truDel2K(
    final delayTime=120)
    "2 minute delay for 2K difference";
```

```
Buildings.Controls.OBC.CDL.Reals.Hysteresis hysValPos95(
  final uLow=0.85,
  final uHigh=0.95)
  "Hysteresis for valve position around 95%";

Buildings.Controls.OBC.CDL.Reals.Hysteresis hysValPosPla(
  final uLow=0.10,
  final uHigh=0.95)
  "Hysteresis for valve position for plant request";

Buildings.Controls.OBC.CDL.Logical.Not not3K
  "Negation of 3K condition";

Buildings.Controls.OBC.CDL.Logical.And and2K
  "2K condition without 3K condition";

Buildings.Controls.OBC.CDL.Logical.Or orTemp
  "Any temperature condition active";

Buildings.Controls.OBC.CDL.Logical.Not notTemp
  "No temperature conditions active";

Buildings.Controls.OBC.CDL.Logical.And andValve
  "Valve condition when no temperature conditions";

Buildings.Controls.OBC.CDL.Integers.Sources.Constant intCon0(
  final k=0)
  "Integer constant 0";

Buildings.Controls.OBC.CDL.Integers.Sources.Constant intCon1(
  final k=1)
  "Integer constant 1";

Buildings.Controls.OBC.CDL.Integers.Sources.Constant intCon2(
  final k=2)
  "Integer constant 2";

Buildings.Controls.OBC.CDL.Integers.Sources.Constant intCon3(
  final k=3)
  "Integer constant 3";

Buildings.Controls.OBC.CDL.Integers.Switch swiChiWat3K
  "Switch for 3K temperature condition";

Buildings.Controls.OBC.CDL.Integers.Switch swiChiWat2K
  "Switch for 2K temperature condition";

Buildings.Controls.OBC.CDL.Integers.Switch swiChiWatValve
  "Switch for valve position condition";

Buildings.Controls.OBC.CDL.Integers.Switch swiChiPla
```

"Switch for chiller plant request";

equation
  connect(TAirSup, sub1.u1);
  connect(TAirSupSet, sub1.u2);
  connect(sub1.y, greThe3K.u);
  connect(sub1.y, greThe2K.u);
  connect(greThe3K.y, truDel3K.u);
  connect(greThe2K.y, truDel2K.u);
  connect(uCooCoi, hysValPos95.u);
  connect(uCooCoi, hysValPosPla.u);

  connect(truDel3K.y, not3K.u);
  connect(not3K.y, and2K.u1);
  connect(truDel2K.y, and2K.u2);

  connect(truDel3K.y, orTemp.u1);
  connect(and2K.y, orTemp.u2);
  connect(orTemp.y, notTemp.u);

  connect(notTemp.y, andValve.u1);
  connect(hysValPos95.y, andValve.u2);

  connect(truDel3K.y, swiChiWat3K.u2);
  connect(intCon3.y, swiChiWat3K.u1);
  connect(intCon0.y, swiChiWat3K.u3);

  connect(and2K.y, swiChiWat2K.u2);
  connect(intCon2.y, swiChiWat2K.u1);
  connect(swiChiWat3K.y, swiChiWat2K.u3);

  connect(andValve.y, swiChiWatValve.u2);
  connect(intCon1.y, swiChiWatValve.u1);
  connect(swiChiWat2K.y, swiChiWatValve.u3);

  connect(swiChiWatValve.y, yChiWatResReq);

  connect(hysValPosPla.y, swiChiPla.u2);
  connect(intCon1.y, swiChiPla.u1);
  connect(intCon0.y, swiChiPla.u3);
  connect(swiChiPla.y, yChiPlaReq);

  annotation (
    defaultComponentName="chiWatPlaReq",
    Icon(coordinateSystem(preserveAspectRatio=false), graphics={
      Rectangle(
        extent={{-100,-100},{100,100}},
        lineColor={0,0,127},
        fillColor={255,255,255},
        fillPattern=FillPattern.Solid),
      Text(

```
    extent={{-150,150},{150,110}},
    textString="%name",
    lineColor={0,0,255})}),
Diagram(coordinateSystem(preserveAspectRatio=false, extent={{-200,-120},{200,120}}, grid={2,2}),
  graphics={
    Rectangle(extent={{-190,110},{-130,40}}, lineColor={28,108,200}, pattern=LinePattern.Dash),
    Rectangle(extent={{-120,110},{-60,40}}, lineColor={28,108,200}, pattern=LinePattern.Dash),
    Rectangle(extent={{-50,110},{10,40}}, lineColor={28,108,200}, pattern=LinePattern.Dash),
    Rectangle(extent={{20,110},{80,40}}, lineColor={28,108,200}, pattern=LinePattern.Dash),
    Rectangle(extent={{90,110},{150,40}}, lineColor={28,108,200}, pattern=LinePattern.Dash),
    Rectangle(extent={{160,110},{190,40}}, lineColor={28,108,200}, pattern=LinePattern.Dash),
    Rectangle(extent={{-120,30},{-60,-40}}, lineColor={28,108,200}, pattern=LinePattern.Dash),
    Rectangle(extent={{160,30},{190,-40}}, lineColor={28,108,200}, pattern=LinePattern.Dash)}));

end Task4;
```

**Task 4 – Human prepared**

```
block PlantRequests "Output plant requests for multizone air handling unit"

  parameter Buildings.Controls.OBC.ASHRAE.G36.Types.HeatingCoil heaCoi=Buildings.Controls.OBC.ASHRAE.G36.Types.HeatingCoil.WaterBased
    "Heating coil type"
    annotation (__cdl(ValueInReference=false));
  parameter Buildings.Controls.OBC.ASHRAE.G36.Types.CoolingCoil cooCoi=Buildings.Controls.OBC.ASHRAE.G36.Types.CoolingCoil.WaterBased
    "Cooling coil type"
    annotation (__cdl(ValueInReference=false));
  parameter Real Thys = 0.1
    "Hysteresis for checking temperature difference"
    annotation(__cdl(ValueInReference=false),
        Dialog(tab="Advanced"));
  parameter Real posHys = 0.05
    "Hysteresis for checking valve position difference"
    annotation(__cdl(ValueInReference=false),
        Dialog(tab="Advanced"));

  Buildings.Controls.OBC.CDL.Interfaces.RealInput TAirSup(
    final unit="K",
    final displayUnit="degC",
    final quantity="ThermodynamicTemperature")
    "Measured supply air temperature"
    annotation (Placement(transformation(extent={{-240,180},{-200,220}}),
      iconTransformation(extent={{-140,60},{-100,100}})));
  Buildings.Controls.OBC.CDL.Interfaces.RealInput TAirSupSet(
    final unit="K",
    final displayUnit="degC",
    final quantity="ThermodynamicTemperature")
    "Setpoint for supply air temperature"
    annotation (Placement(transformation(extent={{-240,140},{-200,180}}),
      iconTransformation(extent={{-140,10},{-100,50}})));
  Buildings.Controls.OBC.CDL.Interfaces.RealInput uCooCoiSet(
```

```
    final unit="1",
    final min=0,
    final max=1) if
      cooCoi == Buildings.Controls.OBC.ASHRAE.G36.Types.CoolingCoil.WaterBased
    "Commanded ooling coil valve position"
    annotation (Placement(transformation(extent={{-240,80},{-200,120}}),
      iconTransformation(extent={{-140,-50},{-100,-10}})));
  Buildings.Controls.OBC.CDL.Interfaces.IntegerOutput yChiWatResReq if
      cooCoi == Buildings.Controls.OBC.ASHRAE.G36.Types.CoolingCoil.WaterBased
    "Chilled water reset request"
    annotation (Placement(transformation(extent={{200,180},{240,220}}),
      iconTransformation(extent={{100,60},{140,100}})));
  Buildings.Controls.OBC.CDL.Interfaces.IntegerOutput yChiPlaReq if
      cooCoi == Buildings.Controls.OBC.ASHRAE.G36.Types.CoolingCoil.WaterBased
    "Chiller plant request"
    annotation (Placement(transformation(extent={{200,0},{240,40}}),
      iconTransformation(extent={{100,10},{140,50}})));

protected
  Buildings.Controls.OBC.CDL.Reals.Subtract cooSupTemDif
    "Find the cooling supply temperature difference to the setpoint"
    annotation (Placement(transformation(extent={{-170,190},{-150,210}})));
  Buildings.Controls.OBC.CDL.Reals.GreaterThreshold greThr(
    final t=3,
    final h=Thys)
    "Check if the supply temperature is greater than the setpoint by a threshold value"
    annotation (Placement(transformation(extent={{-80,190},{-60,210}})));
  Buildings.Controls.OBC.CDL.Reals.GreaterThreshold greThr1(
    final t=2,
    final h=Thys)
    "Check if the supply temperature is greater than the setpoint by a threshold value"
    annotation (Placement(transformation(extent={{-80,140},{-60,160}})));
  Buildings.Controls.OBC.CDL.Logical.TrueDelay truDel(
    final delayTime=120)
    "Check if the input has been true for a certain time"
    annotation (Placement(transformation(extent={{-40,190},{-20,210}})));
  Buildings.Controls.OBC.CDL.Logical.TrueDelay truDel1(
    final delayTime=120)
    "Check if the input has been true for a certain time"
    annotation (Placement(transformation(extent={{-40,140},{-20,160}})));
  Buildings.Controls.OBC.CDL.Reals.GreaterThreshold greThr2(
    final t=0.95,
    final h=posHys) if
      cooCoi == Buildings.Controls.OBC.ASHRAE.G36.Types.CoolingCoil.WaterBased
    "Check if the chilled water valve position is greater than a threshold value"
    annotation (Placement(transformation(extent={{-120,90},{-100,110}})));
  Buildings.Controls.OBC.CDL.Integers.Sources.Constant thr(
    final k=3) "Constant 3"
    annotation (Placement(transformation(extent={{0,222},{20,242}})));
  Buildings.Controls.OBC.CDL.Integers.Switch chiWatRes3 if
      cooCoi == Buildings.Controls.OBC.ASHRAE.G36.Types.CoolingCoil.WaterBased
```

```modelica
    "Send 3 chilled water reset request"
    annotation (Placement(transformation(extent={{160,190},{180,210}})));
  Buildings.Controls.OBC.CDL.Integers.Switch chiWatRes2 if
      cooCoi == Buildings.Controls.OBC.ASHRAE.G36.Types.CoolingCoil.WaterBased
    "Send 2 chilled water reset request"
    annotation (Placement(transformation(extent={{120,140},{140,160}})));
  Buildings.Controls.OBC.CDL.Integers.Sources.Constant two(
    final k=2)
    "Constant 2"
    annotation (Placement(transformation(extent={{0,170},{20,190}})));
  Buildings.Controls.OBC.CDL.Reals.LessThreshold lesThr(
    final t=0.85,
    final h=posHys) if
      cooCoi == Buildings.Controls.OBC.ASHRAE.G36.Types.CoolingCoil.WaterBased
    "Check if the chilled water valve position is less than a threshold value"
    annotation (Placement(transformation(extent={{-120,50},{-100,70}})));
  Buildings.Controls.OBC.CDL.Logical.Latch lat if
      cooCoi == Buildings.Controls.OBC.ASHRAE.G36.Types.CoolingCoil.WaterBased
    "Keep true signal until other condition becomes true"
    annotation (Placement(transformation(extent={{-40,90},{-20,110}})));
  Buildings.Controls.OBC.CDL.Integers.Switch chiWatRes1 if
      cooCoi == Buildings.Controls.OBC.ASHRAE.G36.Types.CoolingCoil.WaterBased
    "Send 1 chilled water reset request"
    annotation (Placement(transformation(extent={{80,90},{100,110}})));
  Buildings.Controls.OBC.CDL.Integers.Sources.Constant one(
    final k=1) "Constant 1"
    annotation (Placement(transformation(extent={{0,110},{20,130}})));
  Buildings.Controls.OBC.CDL.Integers.Sources.Constant zer(
    final k=0) "Constant 0"
    annotation (Placement(transformation(extent={{0,50},{20,70}})));
  Buildings.Controls.OBC.CDL.Logical.Latch lat1 if
      cooCoi == Buildings.Controls.OBC.ASHRAE.G36.Types.CoolingCoil.WaterBased
    "Keep true signal until other condition becomes true"
    annotation (Placement(transformation(extent={{-40,10},{-20,30}})));
  Buildings.Controls.OBC.CDL.Reals.LessThreshold lesThr1(
    final t=0.1,
    final h=posHys) if
      cooCoi == Buildings.Controls.OBC.ASHRAE.G36.Types.CoolingCoil.WaterBased
    "Check if the chilled water valve position is less than a threshold value"
    annotation (Placement(transformation(extent={{-120,4},{-100,24}})));
  Buildings.Controls.OBC.CDL.Integers.Switch intSwi3 if
      cooCoi == Buildings.Controls.OBC.ASHRAE.G36.Types.CoolingCoil.WaterBased
    "Send 1 chiller plant request"
    annotation (Placement(transformation(extent={{80,10},{100,30}})));

equation
  connect(TAirSup, cooSupTemDif.u1) annotation (Line(points={{-220,200},{-180,200},
          {-180,206},{-172,206}}, color={0,0,127}));
  connect(TAirSupSet, cooSupTemDif.u2) annotation (Line(points={{-220,160},{-190,
          160},{-190,194},{-172,194}}, color={0,0,127}));
  connect(cooSupTemDif.y, greThr.u)
```

```
      annotation (Line(points={{-148,200},{-82,200}}, color={0,0,127}));
  connect(greThr.y, truDel.u)
    annotation (Line(points={{-58,200},{-42,200}}, color={255,0,255}));
  connect(greThr1.y, truDel1.u)
    annotation (Line(points={{-58,150},{-42,150}}, color={255,0,255}));
  connect(cooSupTemDif.y, greThr1.u) annotation (Line(points={{-148,200},{-100,200},
          {-100,150},{-82,150}}, color={0,0,127}));
  connect(uCooCoiSet, greThr2.u)
    annotation (Line(points={{-220,100},{-122,100}}, color={0,0,127}));
  connect(truDel.y, chiWatRes3.u2)
    annotation (Line(points={{-18,200},{158,200}}, color={255,0,255}));
  connect(thr.y, chiWatRes3.u1) annotation (Line(points={{22,232},{60,232},{60,208},
          {158,208}}, color={255,127,0}));
  connect(truDel1.y, chiWatRes2.u2)
    annotation (Line(points={{-18,150},{118,150}}, color={255,0,255}));
  connect(two.y, chiWatRes2.u1) annotation (Line(points={{22,180},{50,180},{50,158},
          {118,158}}, color={255,127,0}));
  connect(greThr2.y, lat.u)
    annotation (Line(points={{-98,100},{-42,100}}, color={255,0,255}));
  connect(uCooCoiSet, lesThr.u) annotation (Line(points={{-220,100},{-140,100},{
          -140,60},{-122,60}}, color={0,0,127}));
  connect(lesThr.y, lat.clr) annotation (Line(points={{-98,60},{-60,60},{-60,94},
          {-42,94}}, color={255,0,255}));
  connect(one.y, chiWatRes1.u1) annotation (Line(points={{22,120},{40,120},{40,108},
          {78,108}}, color={255,127,0}));
  connect(lat.y, chiWatRes1.u2)
    annotation (Line(points={{-18,100},{78,100}}, color={255,0,255}));
  connect(chiWatRes1.y, chiWatRes2.u3) annotation (Line(points={{102,100},{110,100},
          {110,142},{118,142}}, color={255,127,0}));
  connect(chiWatRes2.y, chiWatRes3.u3) annotation (Line(points={{142,150},{150,150},
          {150,192},{158,192}}, color={255,127,0}));
  connect(zer.y, chiWatRes1.u3) annotation (Line(points={{22,60},{30,60},{30,92},
          {78,92}}, color={255,127,0}));
  connect(chiWatRes3.y, yChiWatResReq)
    annotation (Line(points={{182,200},{220,200}}, color={255,127,0}));
  connect(greThr2.y, lat1.u) annotation (Line(points={{-98,100},{-80,100},{-80,20},
          {-42,20}}, color={255,0,255}));
  connect(uCooCoiSet, lesThr1.u) annotation (Line(points={{-220,100},{-140,100},
          {-140,14},{-122,14}}, color={0,0,127}));
  connect(lesThr1.y, lat1.clr)
    annotation (Line(points={{-98,14},{-42,14}}, color={255,0,255}));
  connect(lat1.y, intSwi3.u2)
    annotation (Line(points={{-18,20},{78,20}}, color={255,0,255}));
  connect(one.y, intSwi3.u1) annotation (Line(points={{22,120},{40,120},{40,28},
          {78,28}}, color={255,127,0}));
  connect(zer.y, intSwi3.u3) annotation (Line(points={{22,60},{30,60},{30,12},{78,
          12}}, color={255,127,0}));
  connect(intSwi3.y, yChiPlaReq)
    annotation (Line(points={{102,20},{220,20}}, color={255,127,0}));

end PlantRequests;
```